\documentclass[10pt]{article}

\usepackage{setspace}
\usepackage{amsmath}
\usepackage{amsfonts}
\usepackage{amssymb}
\usepackage{graphicx}
\usepackage{microtype}

\begin{document}
\onehalfspacing
\raggedbottom

\title{Uncovering the Specific Product Rule for the Lattice of Questions}
\author{H.R.N.~van~Erp}
\date{}

\maketitle

\begin{abstract}
\noindent We give here the specific product rule for the lattice of questions. This product rule differs from the product rule for the lattice of statements, hence the qualifier `specific'. This is because the elements in the lattice of statements are ordered by way of implication, an upper context, whereas the elements in the lattice of questions are ordered by way of relevancy, a lower context.

\end{abstract}

\section{Introduction}
It is only fitting to quote here the Bayesian forerunner that pointed us all to the importance and fundamental usefulness of the concept of simple ordering, that is, lattice theory, for deriving laws, be they physical or not: 
\\
\begin{quotation}
\noindent In this paper [we] employ lattice theory to describe the structure of the space of assertions and demonstrate how logical implication on the Boolean lattice provides the framework on which the calculus of inductive inference is constructed. [We] then introduce questions by following Cox who defined a question in terms of the set of assertions that can answer it. The lattice structure of questions is then explored and the calculus for manipulating the relevance of a question to an unresolved issue is examined.
\begin{quotation}
	\qquad \textit{What is a Question?}, (Knuth, 2003) 
\end{quotation}
\end{quotation}


\section{What is a Question?}
In his last scientific publication Cox explored the logic of inquiry, \cite{Cox}. In this paper he defined a question as a set of all possible statements that answer it. So, given a hypothesis space of possible answers, that is, statements, one can construct questions, and compare their equivalence by comparing their sets of answers. 

For example, both questions `is it raining?' and `is it not raining?' are equivalent since they are both answered by the same set of statements, \cite{Knuth.3}. Say, we have three possible weather states
\[
	a \equiv \text{rain}, \quad b \equiv \text{snow}, \quad a \equiv \text{sun}
\]
For compactness of notation, let
\[
	A \equiv a, \quad B \equiv b , \quad C \equiv c
\]
and let
\[
	AB \equiv a + b, \quad AC \equiv a + c, \quad BC \equiv b + c, \quad ABC \equiv a + b + c
\]
where `$+$' is the symbol for disjunction and $AB$ is the statement `it is either raining or snowing'\footnote{Or, equivalently, in this specific case, where there are only three possible weather states, the statement `it is not sunny'.}. Then the set of all real questions is, ~\cite{Knuth.3}:

\begin{centering}

		$\left. \right.$
		\\
		$\left\{\textit{ABC}\right.$,	
		\\		
		$\left. \right.$
		\\
		$\textit{AB} \vee \textit{AC} \vee \textit{BC}$,	
		\\
		$\left. \right.$
		\\
		$\textit{AB} \vee \textit{AC}$, $\textit{AB} \vee \textit{BC}$, $\textit{AC} \vee \textit{BC}$,
		\\
		$\left. \right.$
		\\
		$\textit{AB} \vee \textit{C}$, $\textit{AC} \vee \textit{B}$, $\textit{BC} \vee \textit{A}$,
		\\
		$\left. \right.$
		\\
		$\left.\textit{A} \vee \textit{B} \vee \textit{C}\ \right\}$ 
		\\
		$\left. \right.$
	
\end{centering}

\noindent where the symbol `$\vee$' stands for disjunction. 

Note that we use the symbol `$+$' for the construction of the atomic elements of which the questions are constituted, whereas the symbol `$\vee$' is used to combine these atomic elements into real questions. Also note that the questions have been ordered by set inclusion, henceforth denoted as `$\leq$', with the most ambiguous question, that is, the question with the largest set of possible answers, $ABC$, at the top, and the most concise question, that is, the question with the smallest set of possible answers, that is, $A\vee B\vee C$, at the bottom.   

The set of possible answers to the question `is it sunny or not?', that is, $AB\vee C$, constitutes that selfsame question `is it sunny or not?': 
\begin{equation}
\label{1.0a}
	AB\vee C = \left\{a,b,c, a + b\right\}
\end{equation}
and this question actually might, if it is sunny, give us the current weather state, when answered. Alternatively, the question `is it not raining, not snowing, or not sunny?' is answered by all the elements in the set $AB\vee AC\vee BC$, where   
\begin{equation}
\label{1.0b}
	AB\vee AC\vee BC = \left\{a,b,c, a + b, a + c, b + c\right\}
\end{equation}
and this question, when answered, will only manage to exclude one possible weather state, no matter what kind of weather we are in.

So, a question is a set of statements that can be given as an answer to that question. As each question represents a set of answers, related questions may be ordered by set inclusion. This ordering relation of set inclusion implements the concept of answering, \cite{Knuth.3}. So, if question, say, $Q_{1}$ is a subset of question $Q_{2}$, then $Q_{1}\leq Q_{2}$ and by answering question $Q_{1}$ we will have necessarily answered question $Q_{2}$. For example, the question `is it raining, snowing or sunny?', that is,
\begin{equation}
\label{1.1}
 A \vee B \vee C = \left\{ a,b,c\right\} 
\end{equation}
also answers the question `is it raining or not?'
\begin{equation}
\label{1.2}
	A\vee BC = \left\{ a,b,c, b + c\right\}
\end{equation}
Comparing the sets corresponding with questions \eqref{1.1} and \eqref{1.2}, we may easily check that the former is indeed included by the latter
\begin{equation}
\label{1.3}
\left\{ a,b,c\right\} = 	A\vee B\vee C \leq AB\vee C = \left\{ a,b,c, a + b\right\}
\end{equation}

Since questions are just sets of all the possible statements that answer that question, we have that the logical meet, `$\cap$', and join, `$\cup$', of set theory may be applied to questions. The meet of the questions $AB\vee C$, `is it sunny or not?',  and  $A\vee BC$, `is it raining or not?', gives the question `is it raining, snowing, or sunny?':
\begin{align}
\label{1.4}
	\left(AB\vee C\right)\cap \left(A\vee BC\right) &= \left\{ a,b,c, a + b\right\} \cap  \left\{ a,b,c, b + c\right\} \nonumber\\
	\nonumber\\
	&= \left\{ a,b,c\right\} \\
	\nonumber\\
	&= A\vee B\vee C \nonumber
\end{align}
This may be seen as follows. If we first ask if it is sunny, then we will either know that it is sunny or not. If it is not sunny, then we may inquire further and ask whether then it is raining or not. After which we will know exactly what kind of weather it is. We would have gotten the same result had we asked directly whether it was raining, snowing, or sunny. So, the meet of two questions gives us a question that is more informative, when answered, than either question alone. 

The join of the questions $AB\vee C$, `is it sunny or not?', and $A\vee BC$, `is it raining or not?' gives the question `is it not raining or is it not sunny?':
\begin{align}
\label{1.5}
	\left(AB\vee C\right)\cup \left(A\vee BC\right) &= \left\{ a,b,c, a + b\right\} \cup  \left\{ a,b,c, b + c\right\} \nonumber\\
	\nonumber\\
	&= \left\{ a,b,c, a + b, b + c\right\} \\
	\nonumber\\
	&= AB\vee BC \nonumber
\end{align}
We can see that a join of two given questions gives us a less informative question than either of the questions alone, or, for that matter, the meet of those two same questions. This fact will prove to be crucial in the derivation of the specific product rule of the lattice of questions.

\section{Lattice Theory and Quantification}
Two elements of a set are ordered by comparing them according to a binary ordering relation, that is, by way of `$\leq$', which may be read as `is included by'. Elements may be comparable, in which case they form a chain, or they may be incomparable, in which case they form a anti-chain. A set consisting of both inclusion and incomparability are called are called partially ordered sets, or posets for short, \cite{Knuth.5}.

Given a set of elements in a poset, their upper bound is the set of elements that contain them. Given a pair of elements $x$ and $y$, the least element of the upper bound is called the join, denoted  $x\vee y$. The lower bound of a pair of elements is defined dually by considering all the elements that the pair of elements share. The greatest elements of the lower bound is called the meet, denoted $x\wedge y$. Note that we have over-loaded the symbol `$\vee$', which still stands for disjunction, though now in a different, that is, more general, context than previously. 

A lattice is a partially ordered set where each pair of elements has a unique meet and unique join. There often exist elements that are not formed from the join of any pair of elements. These elements are called join-irreducible elements. Meet-irreducible elements are defined similarly. We can choose to view and join and meet as algebraic operations that take any two lattices elements to a unique third lattice element. From this perspective, the lattice is an algebra.

An algebra can be extended to a calculus by defining functions that take lattice elements to real numbers. This enables one to quantify the relationships between the lattice elements. A valuation $v$ is a function that takes a single lattice element $x$ to a real number $v\left(x\right)$ in a way that respects the partial order, so that, depending on the type of algebra, either $v\left(x\right) \leq v\left(y\right)$ or $v\left(y\right) \leq v\left(x\right)$, if in the poset we have that $x\leq y$. This means that the lattice structure imposes constraints on the valuation assignments, which can be expressed as a set of constraint equations, \cite{Knuth.6}.

\subsection{The general sum rule}
In what follows we will closely follow the exposition of \cite{Knuth.6}. This exposition is a beautiful re-telling of a tale already told in Chapter~2 of Jaynes' \cite{Jaynes}. Though Knuth's narration is much more abstract and, consequently, much more general. Point in case being that it will lead us, amongst other things, \cite{Knuth.2}, to the derivation of a truly Bayesian information theory, also called inquiry calculus, \cite{Knuth.3}. It is only when we arrive at the specific product rule of inquiry calculus that we will deviate, ever so slightly, from \cite{Knuth.6}. 

\begin{figure}[!h]
	\centering
		\includegraphics[width=0.30\textwidth]{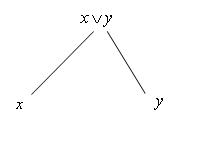}
	\caption{partial lattice of $x\vee y$}
	\label{plot5c}
\end{figure}

We begin by considering a special case of elements $x$ and $y$ with join $x\vee y$ and a null-meet $x\wedge y = \emptyset$. In Figure~\ref{plot5c} we give the graphical representation of this simple lattice, with the exclusion of the bottom element $x\wedge y$, which, being empty, represents an impossibility: 
\\
\\

\noindent The value we assign to the join $x\vee y$, written $v\left(x\vee y\right)$, must be a function of the values we assign to both $x$ and $y$, $v\left(x\right)$ and $v\left(y\right)$. Since if there did not exist any functional relationship, then the valuation could not possibly reflect the underlying lattice structure; that is, valuation must maintain ordering, in the sense that $x \leq x\vee y$ implies either  $v\left(x\right) \leq  v\left(x\vee y\right)$, lattice of statements, or $v\left(x\right) \geq v\left(x\vee y\right)$, lattice of questions. So, we write this functional relationship in terms of an unknown binary operator $\oplus$:
\begin{equation}
	\label{eq36}
		v\left(x\vee y\right) = v\left(x\right)\oplus v\left(y\right)
\end{equation}

\begin{figure}[!h]
	\centering
		\includegraphics[width=0.60\textwidth]{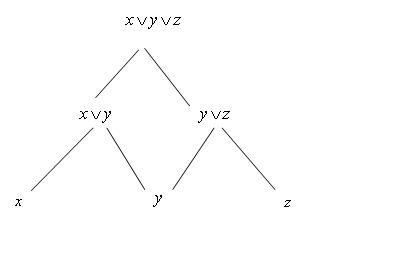}
	\caption{partial lattice of $x\vee y\vee z$}
	\label{plot6c}
\end{figure}

We now consider another case where we have three elements  $x$,  $y$, and  $z$, such that their meets are again disjoint, Figure~\ref{plot6c}:
\\
\\
\\
\\
\\
\\
\\
\\
\\
\\
\\
\\
\\

\noindent Because of the associativity of the join, we have that the least upper bound of these three elements, $x\vee y\vee z$, can be obtained in two different ways: $x\vee \left(y\vee z\right)$ and $\left(x\vee y\right)\vee z$. By applying \eqref{eq36}, the value we assign to this join can also be obtained in two different ways: $v\left(x\right)\oplus \left[ v\left(y\right)\oplus v\left(z\right)\right]$ and $\left[v\left(x\right)\oplus v\left(y\right)\right]  \oplus v\left(z\right)$. Consistency then demands that these equivalent assignments have the same value:
\begin{equation}
	\label{eqTja}
		 v\left(x\right)\oplus \left[ v\left(y\right)\oplus v\left(z\right)\right] = \left[v\left(x\right)\oplus v\left(y\right)\right]  \oplus v\left(z\right)
\end{equation}	
This the functional equation for the operator $\oplus$ for which the general solution is given by~\cite{Aczel}:
\begin{equation}
	\label{eq37}
		 f\left[v\left(x\vee y\right)\right] =f\left[v\left(x\right)\right] +  f\left[v\left(y\right)\right]
\end{equation}	
where $f$ is an arbitrary invertible function, so that many valuations are possible. We define the valuation $u$ as 
\[
u\left(x\right) \equiv f\left[v\left(x\right)\right]
\]
and rewrite \eqref{eq37} as
\begin{equation}
	\label{eq38}
		 u\left(x\vee y\right) = u\left(x\right) +  u\left(y\right)
\end{equation}	

\begin{figure}[!h]
	\centering
		\includegraphics[width=0.60\textwidth]{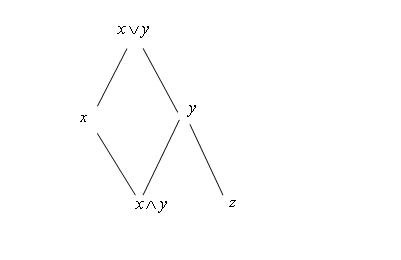}
	\caption{extended partial lattice of $x\vee y$}
	\label{plot7}
\end{figure}

Now that we have a constraint on the valuation for our simple example, we seek the general solution for the entire lattice. To derive the general case, we consider the lattice in Figure \ref{plot7}, where the elements $x\wedge y$ and $z$ have a null meet, as do the elements and $x$ and $z$:
\\
\\

\noindent Applying \eqref{eq38} to these two cases, we get
\begin{equation}
	\label{eq39}
	u\left(y\right) = u\left(x\wedge y\right) + u\left(z\right)
\end{equation}
and, since $y$ is just the join of the part it shares with $x$ joined with $z$,
\begin{equation}
	\label{eq40}
	u\left(x\vee y\right) = u\left(x\right) + u\left(z\right)	
\end{equation}
Substituting for $u\left(z\right)$ in \eqref{eq39} and in \eqref{eq40}, we get the general sum rule:
\begin{equation}
	\label{eqOk}
	u\left(x\vee y\right) = u\left(x\right) + u\left(y\right)	- u\left(x\wedge y\right)
\end{equation}	 	
In general for bi-valuations we have
\begin{equation}
	\label{eq41}
	w\left(\left.x\vee y\right|t\right) = w\left(\left.x\right|t\right) + w\left(\left.y\right|t\right)	- w\left(\left.x\wedge y\right|t\right)
\end{equation}	 	
for any context $t$. Note that the sum rule is not focused solely on joins since it is symmetric with respect to interchange of joins and meets.

At this point Knuth has derived additivity of the measure, which is considered to be an axiom of measure theory. This is significant in that associativity constrains us to have additive measures - there is no other option, \cite{Skilling}.
	
\subsection{Chain rule}	
\begin{figure}[!h]
	\centering
		\includegraphics[width=0.60\textwidth]{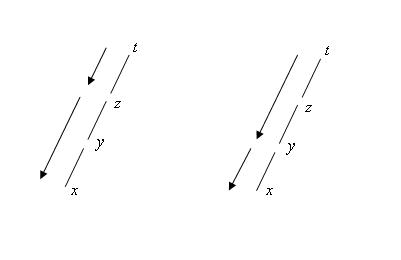}
	\caption{context lattice of $t$}
	\label{plot8}
\end{figure}

We now focus on bi-valuations and explore changes in context~\cite{Knuth.6}. We begin with a special case and consider four ordered elements $x \leq y \leq z \leq t$. The relationship $x\leq z$ can be divided in two relations, $x\leq y$ and $y\leq t$. In the event that $z$ is considered to be the context, this sub-division implies that the context can be considered in parts. Thus the bi-valuation we assign to $x$  with respect to the context $z$, that is, $w\left(\left.x\right|z\right)$, must be related to both the bi-valuation we assign to $x$ with respect to the context $y$, that is, $w\left(\left.x\right|y\right)$, and the bi-valuation we assign to $y$ with respect to the context $z$, that is, $w\left(\left.y\right|z\right)$. That is, there exists a binary operator $\otimes$ that relates the bi-valuations assigned to the two steps to the bi-valuation assigned to the one step, that is,
\begin{equation}
	\label{eq42}
	w\left(\left.x\right|z\right) = w\left(\left.x\right|y\right)\otimes w\left(\left.y\right|z\right)
\end{equation}
By extending this to three steps, Figure~\ref{plot8}:
\\
\\

\noindent and considering the bi-valuation $w\left(\left.x\right|t\right)$, relating $x$ and  $t$, via intermediate contexts $y$ and $z$, results, by way of \eqref{eq42}, in another associativity relationship  	
\begin{equation}
	\label{eq43}
	\left[w\left(\left.x\right|y\right)\otimes w\left(\left.y\right|z\right)\right]\otimes w\left(\left.z\right|t\right) = w\left(\left.x\right|y\right)\otimes\left[ w\left(\left.y\right|z\right)\otimes w\left(\left.z\right|t\right)\right]
\end{equation}	
Using the associativity theorem again results in a constraint equation for non-negative bi-valuations involving changes in context~\cite{Knuth.6}. We call this the chain rule \textit{for an upper context}:
\begin{equation}
	\label{eq44}
	w\left(\left.x\right|z\right) = w\left(\left.x\right|y\right) w\left(\left.y\right|z\right)
\end{equation}	

\begin{figure}[!h]
	\centering
		\includegraphics[width=0.60\textwidth]{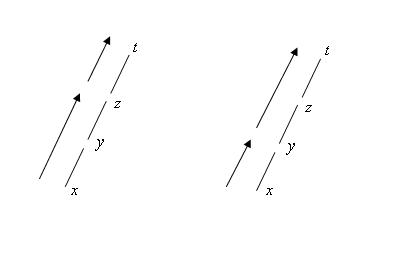}
	\caption{context lattice of $x$}
	\label{plot8b}
\end{figure}

Alternatively, and this is where we generalize on \cite{Knuth.6} in order to arrive later on at the specific product rule of the question of lattices, if $x$ is considered to be the context, rather then $z$, then the sub-division of $x\leq z$ in relations $x\leq y$ and $y\leq t$ also implies that the context can be considered in parts. The bi-valuation we assign to $z$ with respect to the context $x$, that is, $w\left(\left.z\right|x\right)$, must be related to both the bi-valuation we assign to $z$ with respect to the context $x$, that is, $w\left(\left.y\right|x\right)$, and the bi-valuation we assign to $z$ with respect to the context $y$, that is, $w\left(\left.z\right|y\right)$. So, there again exists a binary operator $\otimes$ that relates the bi-valuations assigned to the two steps to the bi-valuation assigned to the one step
\begin{equation}
	\label{eq42b}
	w\left(\left.z\right|x\right) = w\left(\left.y\right|x\right)\otimes w\left(\left.z\right|y\right)
\end{equation}
By extending this to three steps, Figure~\ref{plot8b}:
\\
\\

\noindent and considering the bi-valuation $w\left(\left.t\right|x\right)$, relating $t$ and  $x$, via intermediate contexts $z$ and $y$, results, through \eqref{eq42b}, in the alternative associativity relationship  	
\begin{equation}
	\label{eq43b}
	\left[w\left(\left.y\right|x\right)\otimes w\left(\left.z\right|y\right)\right]\otimes w\left(\left.t\right|z\right) = w\left(\left.y\right|x\right)\otimes\left[ w\left(\left.z\right|y\right)\otimes w\left(\left.t\right|z\right)\right]
\end{equation}	
Using the associativity theorem again results in a constraint equation for non-negative bi-valuations involving changes in context~\cite{Knuth.6}. We call this the chain rule \textit{for a lower context}:
\begin{equation}
	\label{eq44b}
	w\left(\left.z\right|x\right) = w\left(\left.y\right|x\right) w\left(\left.z\right|y\right)
\end{equation}

In this section a valuation calculus has been derived. Associativity of the join gives rise to the sum rule, which is symmetric with respect to interchange of joins and meets. Associativity of changes of context result in two possible chain rules for bi-valuations that dictate how valuations should be manipulated when changing context.

\section{Hypothesis space}
The state space is an enumeration of all the possible states that our system may be in. A given individual may not know precisely which state the system is in, but may have some information that rules out some states, but not others. This set of potential states, then, defines what one can say about the state of the system. For this reason, we call a set of potential states a statement. A statement describes a state of knowledge about the state of the system. The set of all possible statements is called the hypothesis space, \cite{Knuth.1}. 

The lattice of statements is generated by taking the power set, which is the set of all possible subsets of the set of all states, and ordering them according to set inclusion. For a system of n possible states, there are $\sum_{i}^{n}\binom{n}{i} = 2^{n}$ statements including the null set. The bottom element is often omitted from the diagram due to the fact that it represents the logical absurdity. The statement at the top is the truism, generically called the top and denoted `T', which represents the state of knowledge where one only knows that the system can be in one of n possible states. 

The ordering relation of set inclusion naturally encodes logical implication, such that a statement implies all the statements above it. Logical deduction is straightforward in this framework since a statement in the lattice implies (is included by) every element above it with certainty. For example, $x$ implies $x\vee y$ as well as $x\vee y\vee z$, etc. In this sense a lattice is an algebra. 

Logical induction works backwards. One would like to quantify the degree to which one's current state of knowledge implies a statement of greater certainty below it. Since statements do not imply statements below them, this requires a generalization of the algebra representing ordering. In the previous section we have laid the groundwork for generalizing this algebra to a calculus by introducing quantification. In what follows we derive a measure, called probability, that quantifies the degree to which one statement implies another, \cite{Knuth.1}.
	
\subsection{Specific product rule for the lattice of statements}	
\begin{figure}[!h]
	\centering
		\includegraphics[width=0.60\textwidth]{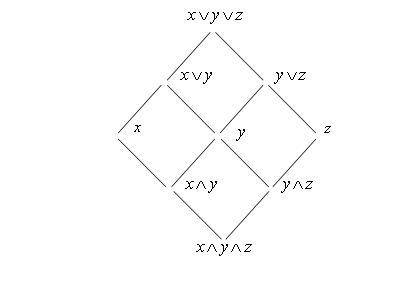}
	\caption{partial lattice of statements}
	\label{plot9}
\end{figure}

We now focus on applying the sum and chain rule to the lattice of statements in Figure~\ref{plot9}, where the elements $x$, $y$, $z$ are understood to be statements:
\\
\\

\begin{figure}[!h]
	\centering
		\includegraphics[width=0.40\textwidth]{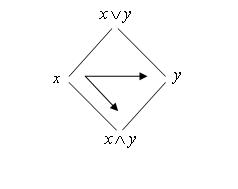}
	\caption{diamond $x$, $x\vee y$, $y$, and $x\wedge y$}
	\label{plot10}
\end{figure}

\noindent To begin we focus on the small diamond in Figure~\ref{plot9} defined by $x$, $x\vee y$, $y$, and $x\wedge y$. If we consider the context to be $x$, then the sum rule \eqref{eq41} may be written down as:
\begin{equation}
	\label{eq45}
	w\left(\left.x\vee y\right|x\right) + w\left(\left.x\wedge y\right|x\right)= w\left(\left.x\right|x\right) + w\left(\left.y\right|x\right)
\end{equation}
Since $x\leq x$ and  $x\leq x\vee y$, we have that the statement $x$ implies both statements $x$ and  $x\vee y$ with absolute certainty, that is, $w\left(\left.x\right|x\right) = w\left(\left.x\vee y\right|x\right) = 1$, and the sum rule \eqref{eq45} implies  	
\begin{equation}
	\label{eq46}
	w\left(\left.x\wedge y\right|x\right) = w\left(\left.y\right|x\right)
\end{equation}
This relationship is expressed by the equivalence of the arrows in Figure~\ref{plot10}:
\\
\\
\\
\\
\\
\\

\begin{figure}[!h]
	\centering
		\includegraphics[width=0.60\textwidth]{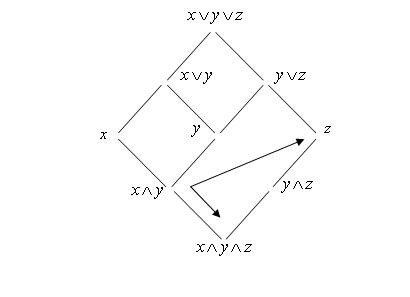}
	\caption{parallelogram  $x\wedge y$, $x\wedge y\wedge z$, $z$, and  $y\vee z$}
	\label{plot11}
\end{figure}

Consider the chain where the bi-valuation $w\left(\left.x\wedge y\wedge z\right|x\right)$ with context $x$ is decomposed into two parts by introducing the intermediate context. The chain rule of an upper context, \eqref{eq44}, gives
\begin{equation}
	\label{eq47}
	w\left(\left.x\wedge y\wedge z\right|x\right) = w\left(\left.x\wedge y\wedge z\right|x\wedge y\right) w\left(\left.x\wedge y\right|x\right)
\end{equation}
To simplify this relation, consider the parallelogram in Figure~\ref{plot9} defined by $x\wedge y$, $x\wedge y\wedge z$, $z$, and  $y\vee z$. From the sum rule \eqref{eq41} with context $x\wedge y$ and the fact that the statement $x\wedge y$ implies both $x\wedge y$ and $y\vee z$ with absolute certainty, that is, $w\left(\left.x\wedge y\right|x\wedge y\right) = w\left(\left.y\vee z\right|x\wedge y\right) = 1$, we obtain
\begin{equation}
	\label{eq48}
	w\left(\left.x\wedge y\wedge z\right|x\wedge y\right) = w\left(\left.z\right|x\wedge y\right)
\end{equation}
This can also be seen by noting that the diamond in Figure~\ref{plot10} has the self-same topology as the right-hand parallelogram in Figure~\ref{plot11}: 
\\
\\
\\
\\
\\
\\
\\
\\
\\
\\
\\
\\

\begin{figure}[!h]
	\centering
		\includegraphics[width=0.60\textwidth]{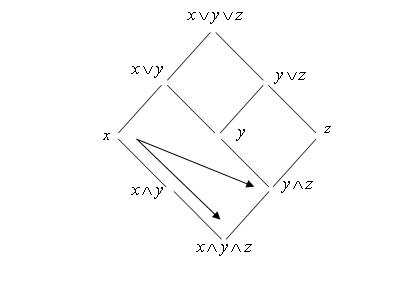}
	\caption{parallelogram $x$, $x\vee y$, $y\wedge z$,  and $x\vee y\vee z$}
	\label{plot12}
\end{figure}

We have simplified the first term on the right hand side of \eqref{eq47}. Now, in order to also simplify the left hand side of \eqref{eq47}, we consider the parallelogram in Figure~\ref{plot9} defined by $x$, $x\vee y$, $y\wedge z$,  and $x\vee y\vee z$. From the sum rule \eqref{eq41} with context $x$ and the fact that the statement $x$ implies both statements $x$ and $x\vee y$ with absolute certainty, that is, $w\left(\left.x\right|x\right) = w\left(\left.x\vee y\right|x\right) = 1$, we have
\begin{equation}
	\label{eq49}
	w\left(\left.x\wedge y\wedge z\right|x\right) = w\left(\left.y\wedge z\right|x\right)
\end{equation}
This can also be seen by noting that the diamond in Figure~\ref{plot10} has the self-same topology as the left-hand parallelogram in Figure~\ref{plot12}: 
\\
\\
\\
\\
\\
\\
\\
\\
\\
\\
\\
\\
\\
\\

\noindent By substituting the simplifications \eqref{eq46}, \eqref{eq48}, and \eqref{eq49} into \eqref{eq47} we obtain the specific product rule the lattice of statements:
\begin{equation}
	\label{eq50}
	w\left(\left.y\wedge z\right|x\right) = w\left(\left.z\right|x\wedge y\right) w\left(\left.y\right|x\right)
\end{equation}

We now have the two necessary rules of probability theory, \eqref{eq45} and  \eqref{eq50}. By relabeling the measure $w$ as $p$, we find the familiar sum and product rule of probability theory:
\begin{align}
	p\left(\left.x\vee y\right|t\right) &= p\left(\left.x\right|t\right) + p\left(\left.y\right|t\right) - w\left(\left.x\wedge y\right|x\right) \\
	p\left(\left.x\wedge y\right|t\right) &= p\left(\left.y\right|x\wedge t\right) p\left(\left.x\right|t\right) \nonumber
\end{align}
where $t$ is typically a context situated higher on the lattice. Now, since for lattices we have that
\[
	a\leq b \Rightarrow a\wedge b = a
\]
we may, for compactness sake, also write the product rule of probability theory as
\begin{equation}
	\label{eq50b}
	p\left(\left.x\wedge y\right|t\right) = p\left(\left.y\right|x\right) p\left(\left.x\right|t\right)
\end{equation}

With respect to probability theory the result of this work by Knuth is a new foundation that encompasses and generalizes both the Cox and Kolmogorov formulations,~\cite{Cox} and~\cite{Kolgomorov}. By introducing probability as a bi-valuation defined on a lattice of statements we can quantify the degree to which one statement implies another. This generalization from logical implication to degrees of implication not only mirrors Cox's notion of plausibility as a degree of belief, but includes it. The main difference is that Cox's formulation is based on a set of desiderata derived from his particular notion of plausibility; whereas here the symmetries of lattices in general form the basis of the theory and the meaning of the derived measure is inherited from the ordering relation, which in the case of statements is implication, \cite{Knuth.6}. 
	
The fact that these lattices are derived from sets means that this works encompasses Kolmogorov's formulation of probability theory as a measure on sets. However, mathematically this theory improves on Kolmogorov's foundation by deriving, rather than assuming, summation. Furthermore, this foundation further extends Kolmogorov's measure-theoretic foundation by introducing the concept of context. This leads directly to probability necessarily being conditional, and Bayes' Theorem follows as a direct result of the chain rule in terms of a change in context, \cite{Knuth.6}.

\section{Inquiry Space}
One can take this idea of states and statements further. By defining a question in terms of the statements that answer it, one can generate the lattice of questions by taking sets of statements as the lattice elements. Just as some statements imply other statements, some questions answer other questions. Answering a given question in the lattice will guarantee that you have answered all the questions above it. One can also generalize this algebra to a calculus by introducing quantification. The result is the inquiry calculus, which is based on a measure called the relevance, that quantifies the degree to which one question answers another, \cite{Knuth.6}. 

\subsection{Specific product rule for the lattice of questions}	
\begin{figure}[!h]
	\centering
		\includegraphics[width=0.60\textwidth]{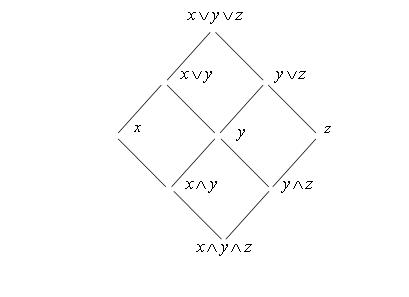}
	\caption{partial lattice of questions}
	\label{plot13}
\end{figure}

We now focus on applying the sum and chain rule to the lattice of questions in Figure~\ref{plot13}, where the elements $x$, $y$, and $z$ are understood to be questions: 
\\
\\
\\
\\
\\
\\

\noindent Answering the meet of questions $x$ and $y$, $x\wedge y$, is equivalent to answering both questions $x$ and $y$ at the same time. Answering the join of questions $x$ and $y$, $x\vee y$, is equivalent to answering either question $x$ or question $y$. See also Section~2. Note that the lattice of statements, Figure~\ref{plot9}, has exactly the same structure as the lattice of questions, Figure~\ref{plot13}. However, they differ in their ordering relations. For the lattice of statements the ordering comes from implication. The higher elements are absolutely implied by the lower elements. In contrast, for the lattice of questions the ordering comes from relevance and the ordering is reversed, in that the lower elements are absolutely relevant for the higher elements. This is the crucial observation that will lead us to the specific product rule for the lattice of questions.

\begin{figure}[!h]
	\centering
		\includegraphics[width=0.40\textwidth]{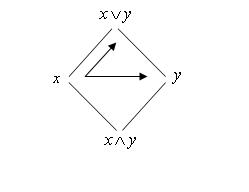}
	\caption{diamond $x$, $x\vee y$, $y$, and $x\wedge y$}
	\label{plot14}
\end{figure}

We again focus on the small diamond defined by $x$, $x\vee y$, $y$, and $x\wedge y$, and we consider the context to be $x$. In the lattice of statements we have that the lower statement implies the higher statement with absolute certainty. In the lattice of questions we have the dual situation that the higher question is answered unambiguously by answering the lower question. So, since  $x\leq x$ and  $x\wedge y\leq x$, we have that question $x$ is answered unambiguously by answering either question $x$ or question $x\wedge y$, that is, $w\left(\left.x\right|x\right) = w\left(\left.x\wedge y\right|x\right) = 1$, and the sum rule \eqref{eq45} implies     
\begin{equation}
	\label{eq51}
	w\left(\left.x\vee y\right|x\right) = w\left(\left.y\right|x\right)
\end{equation}
Comparing \eqref{eq51} with \eqref{eq46}, we see that the reversal in ordering of the elements has led to a corresponding reversal of the `$\wedge$' and the `$\vee$' operators. The relationship \eqref{eq51} is illustrated by the equivalence of the arrows in Figure~\ref{plot14}:
\\
\\

\begin{figure}[!h]
	\centering
		\includegraphics[width=0.60\textwidth]{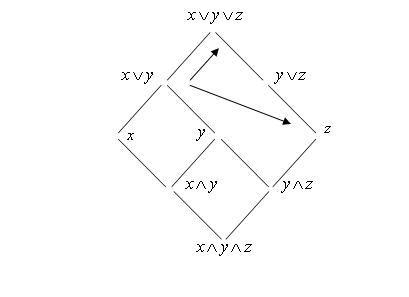}
	\caption{parallelogram $x\vee y$, $x\vee y\vee z$, $z$, and $y\wedge z$}
	\label{plot15}
\end{figure}

\noindent Consider the chain where the bi-valuation $w\left(\left.x\vee y\vee z\right|x\right)$ with context $x$ is decomposed into two parts by introducing the intermediate context. The chain rule of a lower context, \eqref{eq44b}, gives 
\begin{equation}
	\label{eq52}
	w\left(\left.x\vee y\vee z\right|x\right) = w\left(\left.x\vee y\vee z\right|x\vee y\right) w\left(\left.x\vee y\right|x\right)
\end{equation}
To simplify this relation, we consider the parallelogram in Figure~\ref{plot13} defined by $x\vee y$, $x\vee y\vee z$, $z$, and $y\wedge z$. From the sum rule \eqref{eq41} with context $x\vee y$ and the fact that the question $x\vee y$ is answered unambiguously by answering either question $x\vee y$ or question $y\wedge z$, that is, $w\left(\left.x\vee y\right|x\vee y\right) = w\left(\left.y\wedge z\right|x\vee y\right) = 1$, we obtain
\begin{equation}
	\label{eq53}
	w\left(\left.x\vee y\vee z\right|x\vee y\right) = w\left(\left.z\right|x\vee y\right)
\end{equation}
This can also be seen by noting that the diamond in Figure~\ref{plot14} has the self-same topology as the right-hand parallelogram in Figure~\ref{plot15}: 
\\
\\
\\
\\
\\
\\
\\
\\
\\
\\

\begin{figure}[!h]
	\centering
		\includegraphics[width=0.60\textwidth]{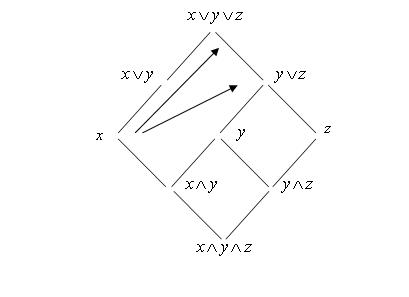}
	\caption{parallelogram $x$, $x\vee y\vee z$, $y\vee z$, and $x\wedge y$}
	\label{plot16}
\end{figure}

We now have simplified the first term on the right hand side of \eqref{eq52}. In order to also simplify the left-hand side of \eqref{eq52}, we consider the parallelogram defined by $x$, $x\vee y\vee z$, $y\vee z$, and $x\wedge y$. From the sum rule \eqref{eq41} with context $x$ and the fact that the question $x$ is answered unambiguously by answering either question $x$ or question $x\wedge y$, that is, $w\left(\left.x\right|x\right) = w\left(\left.x\wedge y\right|x\right) = 1$, we obtain
\begin{equation}
	\label{eq54}
	w\left(\left.x\vee y\vee z\right|x\right) = w\left(\left.y\vee z\right|x\right)
\end{equation}
This can also be seen by noting that the diamond in Figure~\ref{plot14} has the self-same topology as the left-hand parallelogram in Figure~\ref{plot16}: 
\\
\\
\\
\\
\\
\\
\\
\\
\\
\\
\\
\\
\\
\\

\noindent By substituting the simplifications \eqref{eq51}, \eqref{eq53}, and \eqref{eq54} into \eqref{eq52} we obtain the specific product rule for the lattice of question:
\begin{equation}
	\label{eq55}
	w\left(\left.y\vee z\right|x\right) = w\left(\left.z\right|x\vee y\right) w\left(\left.y\right|x\right)
\end{equation}
Note that this product rule for the lattice of questions is just the product rule for the lattice of statements, \eqref{eq50}, with the meet operator `$\wedge$' reversed into a join operator `$\vee$'. This is because of the reversal in ordering, from implication to relevancy.

We now have the two necessary rules of inquiry theory, \eqref{eq45} and  \eqref{eq55}. By relabeling the measure $w$ as $d$, we get the sum and product rule of inquiry calculus:
\begin{align}
	d\left(\left.x\vee y\right|t\right) &= d\left(\left.x\right|t\right) + d\left(\left.y\right|t\right) - d\left(\left.x\wedge y\right|t\right)  \\
	d\left(\left.x\vee y\right|t\right) &= d\left(\left.y\right|x\vee t\right) d\left(\left.x\right|t\right) \nonumber
\end{align}
where $t$ is typically a context situated lower on the lattice. Now, since for lattices we have that
\[
	a\leq b \Rightarrow a\vee b = b
\]
we may, for compactness sake, also write the product rule of inquiry calculus as
\begin{equation}
	\label{eq58}
	d\left(\left.x\vee y\right|t\right) = d\left(\left.y\right|x\right) d\left(\left.x\right|t\right) 
\end{equation}

With respect to inquiry calculus the result of this work is that a whole new field of inference has been opened up. By introducing relevance as a bi-valuation defined on a lattice of questions, \cite{Knuth.6}, we can quantify the degree to which one question is relevant to another. The symmetries of lattices in general form the basis of the theory and the meaning of the derived relevance measure is inherited from the ordering relation, which in the case of questions is relevancy. Because of the concept of context, we have that relevance is necessarily conditional, and a Bayes' Theorem for inquiry theory follows as a direct result of the chain rule in terms of a change in context. 

\section{Discussion}
We have here uncovered the specific product rule for the lattice of questions. We use here the word `uncover' because once the general sum rule and the chain rule for a lower context are in play, then the specific product rule for the lattice of questions follows trivially. 

The specific product rule for the question of lattices was found by this author somewhere in 2009, by way of simple trial and error. But trial and error does not constitute a formal proof. And this author, lacking the necessary mathematical sophistication, had no clue whatsoever on how to proof his conjecture. Only after reading \cite{Knuth.6}, in which the concept of a chain rule is derived, did we have the necessary tools to proof the correctness of the specific product rule of the lattice of questions. Once one adjusts the pertinent equations\footnote{For example, adjust eqs. \{18\} and \{20\} in \cite{Knuth.6} into our eqs. \eqref{eq44b} and \eqref{eq51}, etc.}, then the proof for the specific product rule of the lattice of questions just flows forth naturally from the derivations. 

Note that in \cite{Center} the specific product rule for the lattice of questions, as given here, is also derived. But this proof is more opaque and less accessible than the proof which follows directly from the exceedingly elegant framework given in \cite{Knuth.6}.

\section{Acknowledgments}
We would like to thank Kevin Knuth for the lattice theoretic edifice he has so generously shared with all, and for his kind encouragement and support. Furthermore, we would like to acknowledge that this research has received funding from the European Community's Seventh Framework Programme [FP7/2007-2013] under grant agreement no.~265138.

\end{document}